# Scintillation reduction for combined Gaussian-vortex beam propagating through turbulent atmosphere

G.P. Berman* [a], V.N. Gorshkov[a,b,c], S.V. Torous[b]

[a]Theoretical Division, T-4 & CNLS MS B213, Los Alamos National Laboratory, Los Alamos, New Mexico 87545
[b]National Technical University of Ukraine "KPI," 37 Peremogy Avenue, Building 7, Kiev-56, 03056 Ukraine
[c]The Institute of Physics, National Academy of Sciences of Ukraine, Nauky Ave. 46, Kiev 680028, Ukraine

## ABSTRACT

We numerically examine the spatial evolution of the structure of coherent and partially coherent laser beams (PCBs), including the optical vortices, propagating in turbulent atmospheres. The influence of beam fragmentation and wandering relative to the axis of propagation (z-axis) on the value of the scintillation index (SI) of the signal at the detector is analyzed. A method for significantly reducing the SI, by averaging the signal at the detector over a set of PCBs, is described. This novel method is to generate the PCBs by combining two laser beams - Gaussian and vortex beams, with different frequencies (the difference between these two frequencies being significantly smaller than the frequencies themselves). In this case, the SI is effectively suppressed without any high-frequency modulators.

**Keywords:** atmospheric turbulence, partially coherent beam, scintillation index

**PACS**: 42.25.Dd; 42.25.Fx; 42.60.By; 42.60.Jf; 42.60.By

## 1. INTRODUCTION

The inhomogeneity of the refraction coefficient, $n'(\vec{r})$, in turbulent atmosphere creates fluctuations of the integral intensity at the detector,

$$I^l(z) = \frac{1}{S}\int_S I^l(x,y,z)dxdy, \tag{1}$$

for gigabit data-rates and long-distance optical communications [1,2]. Here $I^l(x,y,z)$ is the intensity of the light field and $S$ is the detector area; the superscript, $l$, indicates a particular state of the atmospheric turbulence. The variation of the signal, $I^l(x,y,z)$, caused by the different spatial distributions of $n'(x,y,z)$, could lead to significant errors when decoding the bits of the detected information. The minimization these errors requires methods that maximally reduce the scintillation index (SI),

$$\sigma^2(z) = \left\langle \left(I^l(z)\right)^2 \right\rangle_l \Big/ \left\langle I^l(z) \right\rangle_l^2 - 1 \equiv \left\langle \left(\delta I^l(z)\right)^2 \right\rangle_l \Big/ \left\langle I^l(z) \right\rangle_l^2. \tag{2}$$

Here the subscript, $l$, indicates an average over many atmospheric states, and $\delta I^l(z) = I^l(z) - \left\langle I^l(z) \right\rangle_l$. The SI, $\sigma^2(z)$, depends on the optical strength of turbulence described by the index of turbulence, $C_n^2$, which is associated with the structure function, $D(r)$, for fluctuations of the refraction index, $n'(\vec{r})$ [3]:

$$D(r) = \left\langle \left[n'(\vec{r}'+\vec{r}) - n'(\vec{r}')\right]^2 \right\rangle_l = C_n^2 r^{2/3}. \tag{3}$$

---

gpb@lanl.gov

Both beam fragmentation and beam wandering (see[4,5] and references therein) are contributors to the SI. Fragmentation is the decay of the initial laser beam into many spatially separated beams. (See Fig. 1*a*.) These separated beams cannot be detected if the detector size is smaller than the characteristic distance between the beams. Wandering is the deviation of the coherent laser beam (initially oriented along the *z*-axis) in the x*y*-plane. This deviation, which depends on *l*, is defined by:

$$\vec{r}_w^l(z) = \int \vec{r} I^l(x,y,z) dx dy \Big/ \int I^l(x,y,z) dx dy . \quad (4)$$

These random deviations of the beam, $\vec{r}_w^l(z)$, can lead to significant random fluctuations of the intensity, $I^l(z)$, and, consequently, to large values of the SI.

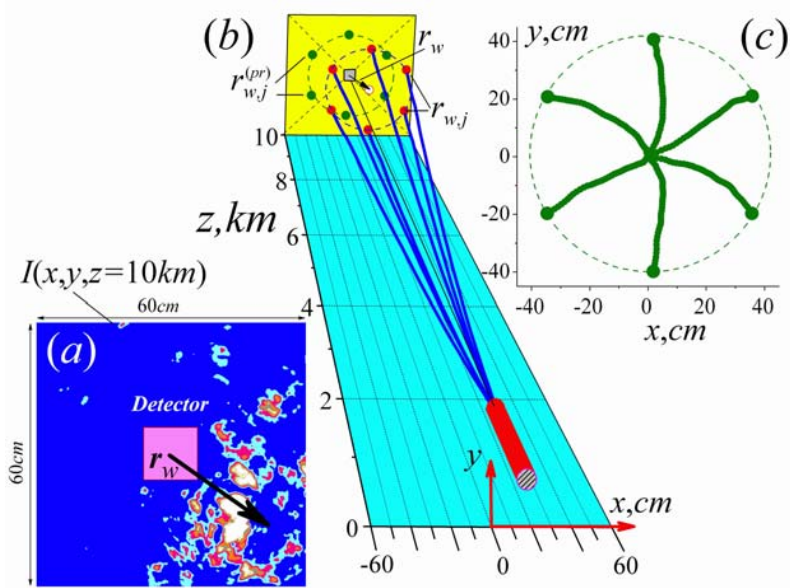

Figure 1. Numerical simulation of the propagation of an initially coherent laser beam through a single realization of a turbulent atmosphere ($\lambda = 1.55 \mu m$; initial laser beam radius, $r_0 = 2cm$, $I(x,y,z=0) \sim \exp(-2r^2/r_0^2)$; $C_n^2 = 2.5 \times 10^{-14} m^{-2/3}$). Note the fragmentation at the detector and the beam wandering. *a*- the distribution of the beam intensity at the plane $z = 10km$ (the initial direction is along the *z*-axis); *b,c* – trajectories of beams in the space and at the $xy$ – plane for different initial directions of the beams relative to the $z$-axis (see text below). The results presented here and below are obtained with the numerical scheme used by us in [6].

It is known that a partially coherent beam (PCB) in combination with a slow time response detector leads to a significant reduction of the SI. (See[7-12] and references therein.) The idea of the method is the following. For a given state of the atmosphere, *l*, a coherent laser beam passes through a phase modulator (PM), which randomly changes the phase of the beam, $\varphi_m(x,y)$ $(m=1,2,...,M)$ in the aperture plane (*m* is the state of the PM with a corresponding value of the intensity, $I_m^l$, at the detector; *M* is the number of realizations integrated by the detector). The signal averaged over *M* PM states at the detector, $\hat{I}^l = \frac{1}{M} \sum_{m=1}^{M} I_m^l$, can become quite stable—almost independent of the atmospheric realizations, *l*. Namely, $\hat{\sigma}^2 \ll \sigma^2$, where $\hat{\sigma}^2$ is the SI calculated by averaging $\hat{I}^l$ over *l*.

The validity of this approach was experimentally demonstrated in[11] and[13] for a simplified model in which the atmospheric turbulence was simulated by a spatial light modulator (atmospheric modulator, AM). A proper choice of PM allowed the authors of [13] to reduce the SI by a factor of 16 (for $M = 10$). This SI reduction is caused by the formation of significantly smaller-scale speckle structures after the PCB passes the AM, compared with the similar situation for a coherent beam. However, the optimistic theoretical predictions were done in[7,8,9] only for the case of strong turbulence and long propagation distances ($z > L_{thres}$, where $L_{thres}$ is the threshold used in the theory based on an asymptotic method). In the region where $\sigma^2(z)$ approaches its maximum values, the theory[7, 8, 9] is not applicable ($z < L_{thres}$).

In this paper, we analyze the case of moderate laser propagation distances, $z < L_{thres}$, and discuss how beam fragmentation and wandering affect the SI, $\sigma^2$. Our main objective is to create algorithms for the PM that allow us to significantly reduce SI for a relatively small number of realizations, $M$. The latter is important for practical implementation of long-distance optical communications. In the process of designing of these algorithms, we have examined the propagation of optical vortices (OV)[14] in turbulent atmospheres. As described in detail below, we recommend the use of a PCB that combines an OV with a Gaussian beam (GB). If the frequencies of these types of beams differ by $\delta\omega$, then for a stationary distribution of the phase mask, $\varphi_0(x,y)$, a significant reduction of the SI can be achieved for time-averaged scintillations of intensity, $\hat{I}^l = \frac{1}{T}\int_0^T I^l(t)dt$, where $T = 2\pi/\delta\omega$.

## 2. SPATIAL EVOLUTION OF A COHERENT LASER BEAM IN A TURBULENT ATMOSPHERE

The reduction of the SI requires investigation of the properties of laser beam spatial evolution while propagating through the atmosphere. It is essential to understand the contributions of wandering and fragmentation to the signal scintillations at the detector, for a variety of levels of atmospheric turbulence, $C_n^2$.

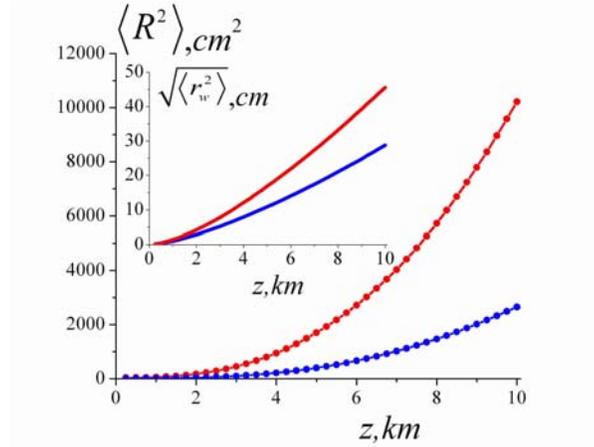

Figure 2. Radius squared and wandering of a coherent beam ($r_0 = 2cm$) as a function of the distance, $z$. Red - $C_n^2 = 10^{-13} m^{-2/3}$; blue - $C_n^2 = 2.5 \times 10^{-14} m^{-2/3}$.

Fig. 2 presents the statistical geometric characteristics of the Gaussian beam with an initial radius, $r_0 = 2cm$, which propagates through a turbulent atmosphere. The root-mean-square radius, $\sqrt{\langle R^2 \rangle}$, characterizes the size of the optical field:

$$\langle R^2(z) \rangle = \left\langle \int r^2 I(x,y,z)dxdy \Big/ \int I(x,y,z)dxdy \right\rangle_l. \tag{5}$$

Note, in Fig. 2, that as the strength of turbulence increases, the wandering of the beam, $\sqrt{\langle r_w^2 \rangle} \equiv \sqrt{\langle (\vec{r}_w^l)^2 \rangle_l}$, relative to its radius, $\sqrt{\langle R^2 \rangle}$, decreases: $\eta = \sqrt{\langle r_w^2 \rangle / \langle R^2 \rangle} = 0.56$ and $0.45$ for $C_n^2 = 2.5 \times 10^{-14} m^{-2/3}$ and for $10^{-13} m^{-2/3}$, correspondingly. The value of the beam radius does not characterize its fragmentation. The measure of fragmentation is defined by the parameter, $\sigma_{I(x,y)}^2(z)$, which is calculated as follows. For the atmospheric state, $l$, the value, $\sigma_{I(x,y),l}^2$, is calculated for a cross section at a given $z$:

$$\sigma_{I(x,y),l}^2 = \frac{\langle (I^l(x,y))^2 \rangle_{x,y} - \langle I^l(x,y) \rangle_{x,y}^2}{\langle I^l(x,y) \rangle_{x,y}^2}. \qquad (6)$$

This averaging is performed over the circle in the $xy$-plane that contains 95% of the total beam power. The center of this circle is located at $\vec{r}_w^l(z)$. Then, these results are averaged over many atmospheric realizations:

$$\sigma_{I(x,y)}^2 = \langle \sigma_{I(x,y),l}^2 \rangle_l . \qquad (7)$$

Larger values of $\sigma_{I(x,y)}^l$ correspond to more significant beam fragmentation. Qualitatively this statement can be interpreted in the following way. For a *fixed size* transverse beam cross-section, the integral intensity can be presented by different numbers of fragments (bright spots). The case of a high degree of fragmentation corresponds to a small number of fragments with high intensity in each of them. This distribution results in a large value of $\langle (I^l(x,y))^2 \rangle_{x,y}$ in (6) for a given value $\langle I^l(x,y) \rangle$. As a result, $\sigma_{I(x,y),l}^2$ is large. In the case of a small-scale field (large number of fragments with small intensities) $\sigma_{I(x,y),l}^2$ is small. Note that the parameter of fragmentation, $\sigma_{I(x,y)}^2$, is averaged over the total cross-section of the beam. In fact, the level of fragmentation is inhomogeneous at the cross-section. A laser beam undergoing amplitude-phase disturbances experiences diffraction in such a way that at the periphery of the beam the speckle structure is mostly small-scale compared with the central part. (See, for example,[15]). This property of the light field will be taken into account in the analysis and interpretation of the results obtained below.

In addition to the results presented in Fig. 2, we calculated the SI, $\sigma^2(z)$, and the level of fragmentation, $\sigma_{I(x,y)}^2$, as a function of the distance, $0 < z \leq 10 km$, for two values of the turbulence strength ($C_n^2 = 2.5 \times 10^{-14} m^{-2/3}$ and $10^{-13} m^{-2/3}$) and for two detector areas ($S = 1 cm^2$ and $100 cm^2$). (See Fig. 3.) The SI, $\sigma^2(z)$, was derived by two methods. In the first case, the center of the detector always is positioned on the $z$-axis. In the second case, for each atmospheric state, $l$, the center of detector is placed at $\vec{r}_w^l(z)$, to the center of the wandering beam. These calculations were made only to analyze the spatial evolution of the beam. This specific SI, which corresponds to the second method of calculation, we denote as $\sigma_{(w)}^2(z)$.

The results presented in Fig. 3 demonstrate that the behavior of $\sigma_{(w)}^2(z)$ is in good agreement with the dependence $\sigma_{I(x,y)}^2(z)$. Namely, the SI of the signal at the "wandering" detector is determined by the beam fragmentation. For the case of weak turbulence (Fig. 3a), the beam wandering evidently dominates at large distances: the SI of the detector fixed at the origin of the $xy$-plane, $\sigma^2(z)$, increases in spite of the decreasing $\sigma_{I(x,y)}^2(z)$. The case of strong turbulence (Fig. 3b) is characterized by relatively weak fragmentation and its sharp decrease in the region $z > 5km$. At first sight it unexpectedly appears that the SI of the wandering detector (the position of which is fixed at $\vec{r}_w^l(z)$) exceeds the SI of the fixed detector for $S = 1 cm^2$; $\sigma_{(w)}^2(z > 5km) > \sigma^2(z > 5km)$. This is related to the above fact that beam fragmentation is inhomogeneous over the cross section of the beam. In the vicinity of the *real axis* of the beam, $\vec{r}_w(z)$, the fragmentation is maximal in contrast to the region $x \approx 0, y \approx 0$, which becomes the periphery of the beam ($\sqrt{\langle r_w^2(z>5km)\rangle} > 20cm$ - see Fig. 2), where the fragmentation is maximal, and for which $\sigma^2(z)$ is calculated. Just because of this, we have: $\sigma^2_{(w)}(z>5km) > \sigma^2(z>5km)$. Note that the beam wandering remains a significant factor in the spatial evolution of the SI, $\sigma^2(z>5km)$: the decrease of SI with increasing $z>5km$ is evidently weaker than the decrease of the fragmentation level, $\sigma^2_{I(x,y)}(z)$, averaged over the beam cross-section. Below we describe a more reliable indication of effect of beam wandering on the SI.

As it pointed out in the introduction, the use of a phase mask which produces small-scale random phase fluctuations over the beam cross section can lead to the result: $\hat{\sigma}^2(z) << \sigma^2(z)$, for the signal,

$$\hat{I}^l = \frac{1}{M}\sum_{m=1}^{M} I_m^l, \tag{8}$$

averaged over the states, $m$, of the PM.

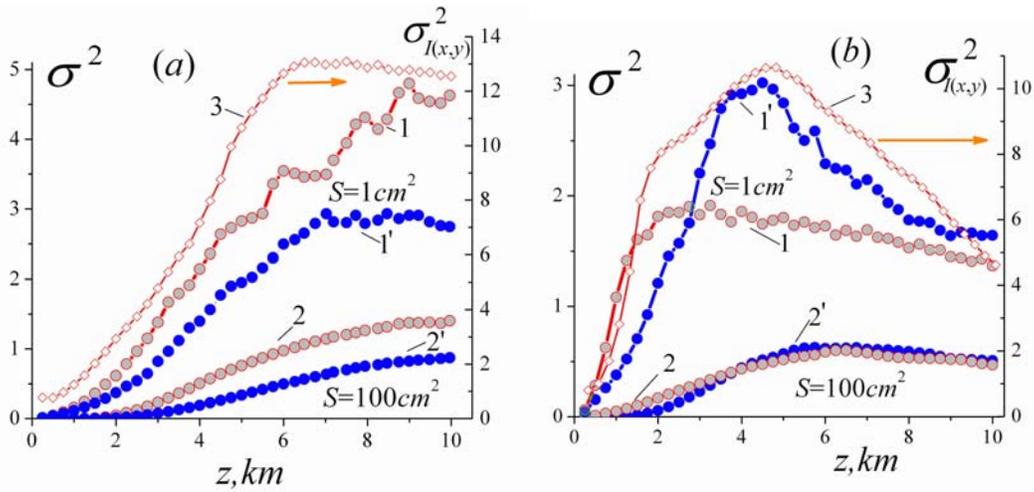

Figure 3. Characteristics of the spatial evolution of the structure of Gaussian beam in turbulent atmosphere. Curves 1, 2 - $\sigma^2(z)$ for detector with area $S = 1cm^2$ и $100cm^2$; $1', 2'$ (right scale)- $\sigma^2_{(w)}(z)$ for detectors of the same size; 3- level of fragmentation $\sigma^2_{I(x,y)}(z)$. a - $C_n^2 = 2.5\times10^{-14} m^{-2/3}$, b- $10^{-13} m^{-2/3}$.

This method was especially effective in the laboratory experiments[13]. But in this case, the beam wandering after the phase modulator, which simulated the atmosphere, was practically absent. Physical atmospheres introduce a significant difference in the spatial dynamics of both types of beams (coherent and PCBs). When wandering develops, the use of a small-scale phase mask is not effective, as shown in our paper [6]. The averaging in (8) is only effective for cases in which the individual signals, $I_m^l(z)$, are statistically independent. This condition is not realized if the PM does not change the initial direction of the beam (the variant of the random phase modulation over the beam cross section). The reason for this is that the deviation of the beams from their initial $z$-direction is mainly due to large-scale inhomogeneities of the index of refraction, $n'$, with characteristic scales, $\Lambda$, that are much larger than the effective beam radius, $\langle R_s \rangle = \sqrt{\langle R^2 \rangle - \langle r_w^2 \rangle}$. Because of this, for a given atmospheric state, $l$, the trajectories of the PCBs and the coherent beams directed along the $z$-axis, are strongly correlated. If the coherent beam deviates away from the detector, the PCB trajectories $\vec{r}_{w,m}^l(z)$ trace, within certain limits, the trajectory $\vec{r}_w^l(z)$. Thus, the averaging (8) does not compensate for beam wandering. However, the small-scale PMs considered can be effective for high level of turbulence, and for long distances from the transmitter when the parameter, $\eta = \sqrt{\langle r_w^2 \rangle / \langle R^2 \rangle}$, is small. For the signal (8), compensation of the beam wandering and significant reduction of SI, $\hat{\sigma}^2(z)$, can only be realized for PCBs with varying directions of the initial propagation. The parameters of these directions must be coordinated with the characteristic parameter $\langle r_w^2(z) \rangle$ in the region of the detector.

## 3. SUPPRESSION OF SCINTILLATIONS BY USING AN ASSYMETRIC OPTICAL VORTEX

We will now describe the basic principles of our method for suppressing the SI. For this, we analyze the case in which a set of PCBs determining the averaged signal (8), is created by the phase mask:

$$\varphi_m(x, y, z = 0) = a_{x,m} x + a_{y,m} y \equiv \vec{a}_m \cdot \vec{r} . \tag{9}$$

Here the modulus of the vectors $\{\vec{a}_m\}, m = 1, 2, ..., M$ is a constant: $|\vec{a}_m| = a_0$.

These phase masks transform the flat wave front of a Gaussian beam, $z = 0$, into the plane, $zk + a_{x,m} x + a_{y,m} y = 0$, and the beam propagates along the normal to the plane, $\vec{n}$, which has components

$$\frac{a_{x,m}}{\sqrt{a_0^2 + k^2}}, \frac{a_{y,m}}{\sqrt{a_0^2 + k^2}}, \frac{k}{\sqrt{a_0^2 + k^2}} . \tag{10}$$

These beams (in fact, coherent beams) are oriented at an angle, $\theta \approx a_0 / k = tg\theta$, to the direction of propagation, $z$, and they are homogeneously distributed in the azimuthal angle, $\beta$. The angle, $\beta$, takes discrete values, $\beta_m = \Delta\beta \cdot m$, $\Delta\beta = \frac{2\pi}{M}$. If the atmosphere is not turbulent, then for each angle, $\beta_j$, one can observe a light spot on the screen positioned perpendicular to the $z$-axis. As the angle, $\beta$, changes, the center of this spot moves discretely along the circumference of a circle with radius, $\rho \approx z\theta$. We assume that the interval of these time series, $\tau$, is much smaller than the characteristic fluctuation time of the atmospheric state, $\tau_{atm} \sim 10^{-3}$ sec : $\tau \ll \tau_{atm}$. We can also call these laser beams PCBs because the variation of direction of coherent beam corresponds to changing its phase (9) at the plane $z = 0$.

Generally, the choice of the "scattering parameter", $\rho$, (angle $\theta$) is not universal, and it must be coordinated with the position of the detector, $z_d$. The results of our simulations [6] demonstrate that the SI, $\hat{\sigma}^2(z_d)$, for the averaged signal (8) can be significantly smaller than the SI, $\sigma^2(z_d)$, if the angle $\theta$ is of the order $\theta_{shot}(z_d) = r_{w,\max}(z_d) / z_d$, or slightly larger. $r_{w,\max}(z)$ is the maximum of the probability density distribution function, $p(r_w)$ (for a given distance, $z$, and a given turbulence strength, $C_n^2$). A judicious choice of aiming, $\theta_{shot}(z_d)$, and the number of PCBs, $M$, in the time series, $\tau$, results in the ratio

$$\hat{\sigma}^2(z_d) = \sigma^2(z_d) / M . \tag{11}$$

(11) is only valid if, for a given atmospheric state, $l$, the values, $I_m^l(z_d)$, are statistically independent. In each case, an optimal number, $M_{sat}$, exists for which the dependence, $\hat{\sigma}^2(z_d, M)$, is saturated for $M > M_{sat}$. In other words, for large $M$ (small angles of rotation, $\Delta\beta$), correlations between values, $I_m^l(z_d)$, develop, and ratio (11) is violated: $\sigma^2(z_d) / \hat{\sigma}^2(z_d) < M$.

A result that illustrates the effectiveness of this approach for suppressing SI is presented in Fig. 4. The angle, $\theta_{shot}$, was chosen so that for $z_d = D = 10 km$, the "scattering parameter", $\rho = \theta_{shot} \times D = 35 cm$  ($r_{w,\max}(D) \approx 25 cm$ if $C_n^2 = 2.5 \times 10^{-14} m^{-2/3}$). A significant reduction of the SI is achieved not only for $z_d = D$, but also for $z < D$. This results from the dependence $\sqrt{\langle r_w^2(z) \rangle}$ being approximately linear. (See Fig. 2).

Note, that for a given detector distance, $z_d$, and turbulence strength, $C_n^2$, the problem must be solved for the optimal choice of both the beam radius, $r_0$, and the "scattering parameter", $\rho(z_d, C_n^2)$, in order to achieve the largest value for $M$, $M_{sat}$.

In Fig. 1b, the characteristic spatial dynamics of the PCBs trajectories, $\vec{r}_{w,m}(z)$, is shown for a single atmospheric state, $l$. Note that the vectors, $\vec{r}_{w,m}(z)$, are distributed on a circle whose center is near the center of the coherent beam, $\vec{r}_w(z)$, initially oriented along the $z$-axis. The reason for this phenomenon is that the deviation of the beams from their initial $z$-directions is mainly due to large-scale inhomogeneities of the index of refraction, $n'$, with characteristic scales, $\Lambda$, that are much larger than the effective beam radius, $R_s(z) = \sqrt{R^2(z) - r_w^2(z)}$, and scattering parameter, $\rho$, for $z \leq 10 km$. So, the trajectories, $\vec{r}_{w,m}(z)$ and $\vec{r}_w(z)$, are strongly correlated. This property of the beams can be used to significantly reduce the SI, if one monitors the center of the beam, $\vec{r}_w^l(z)$, for each atmospheric state, $l$. Then, by predicting the deviation of the leading center of the PCBs, one can target initial directions at the points, $\vec{r}_{shot,m}^l = -\vec{r}_w^l(z_d) + \dfrac{\vec{a}_m}{a_0}\rho(z_d)$, on the plane, $z = z_d$, where $a_0 = k \times \rho(z_d)/z_d$. Under this correction of initial directions, the PCB centers, $\vec{r}_{w,m}^{(pr),l}(z_d)$, will be positioned approximately on a circumference that encloses the detector. (See Fig. 1b; in Fig. 1c, an example of trajectories, $\vec{r}_{w,m}^{(pr),l}(z \leq D)$, is demonstrated on the $xy$-plane for $\rho(z_d = D) = 40 cm$). The results of utilizing this tracking are presented in Fig. 8a, curve 3.

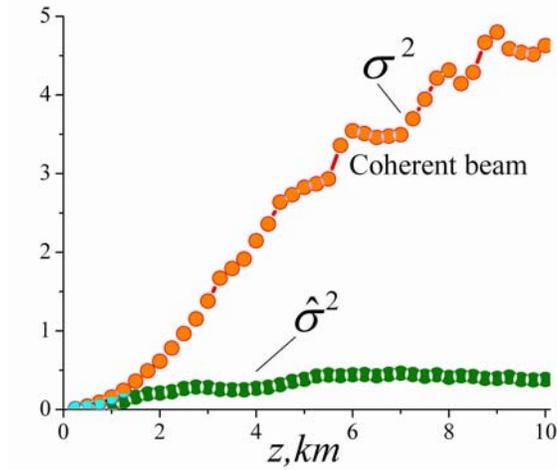

Figure 4. Dependence of the scintillation indexes, $\sigma^2(z)$, $\hat{\sigma}^2(z)$ on the distance, $z$, for detectors with area $S = 1 cm^2$. $C_n^2 = 2.5 \times 10^{-14} m^{-2/3}$, $M = 12$; to the chosen angle, $\theta$, corresponds a "scattering parameter" $\rho = D \times \theta = 35 cm$, $D = 10 km$; $\hat{\sigma}^2(D) \approx \sigma^2(D)/12$. Details of the physical mechanisms responsible for compensation of beam wandering and reduction of the SI, $\hat{\sigma}^2(z)$, are described in [6].

The generation of an optimal set of PCBs that effectively minimizes the SI is a challenging technical problem for high-data-rate optical communication, because the transmission of information through a single frequency channel at rates above $1 Gbit/s$ requires a phase modulator, $\varphi_m(x,y)$, whose frequency above $10^{10} Hz$. Now we discuss our method to eliminate this technical problem.

Consider the properties of a specific PCB that is a superposition of an optical vortex (OV) and a Gaussian beam. The initial directions of these two beams coincide with the $z$-axis. The amplitudes of the OV and the Gaussian beam at the plane $z=0$ are

$$U_{ov}(r,0) = A \times \frac{r}{r_{v0}} \exp\left(-\frac{r^2}{r_{v0}^2}\right) e^{i\beta + i\omega_v t} \text{ and } U_g(r,0) = B \times \exp\left(-\frac{r^2}{r_{g0}^2}\right) e^{i\omega_g t}. \quad (12)$$

For convenience, we present some known results for OVs [14]. One way to create an OV is to pass a Gaussian laser beam through a spiral phase mask [16, 17, 18] which modulates the phase of the beam, $\varphi(x,y) = Arc\tan(y/x) \equiv \beta$, where $\beta$ is the angle of rotation around the direction of propagation, $z$-axis. After passing through the mask, the optical field has the complex amplitude, $U(r, z=0) \sim r\exp(-r^2/r_{v0}^2)e^{i\beta}$. In general, $\varphi(x,y) = \pm j\beta$; where $j$ is an integer that is called the topological charge. Under a paraxial approximation, the spatial evolution of the OV amplitude is described by:

$$U(r,\beta,z) = \frac{A}{w(\tilde{z})^2} \frac{r}{r_{v0}} \exp\left(-\frac{r^2}{r_{v0}^2 \cdot w(\tilde{z})^2}\right) \exp[i\xi(r,\beta,\tilde{z})], \quad (13)$$

where $\tilde{z} = z/z_R$, $z_R = \pi r_{v0}^2/\lambda$, $w(\tilde{z}) = \sqrt{1+\tilde{z}^2}$, $\xi(\rho,\beta,\tau) = 2\arctan\tilde{z} - \tilde{z}\frac{r^2}{r_{v0}^2 \cdot w(\tilde{z})^2} + \beta$.

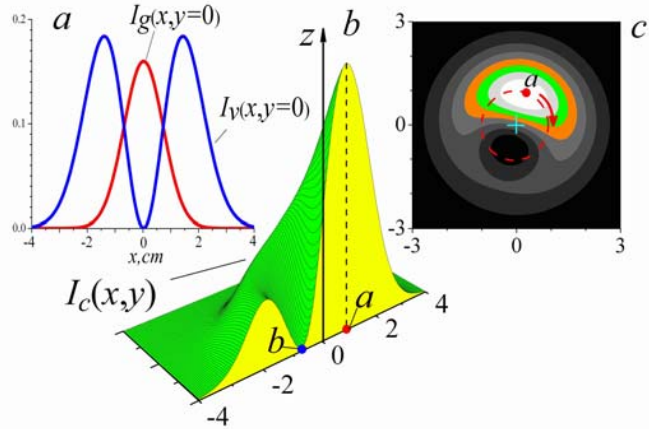

Figure 5. $a$ – Distribution of intensity of an OV, $I_v(x,y)$, and of a Gaussian beam, $I_g(x,y)$, in the plane, $z=0$. $A=1$, $B=0.4$, $r_{v0} = 2cm$, and $r_{g0} = r_{v0}/\sqrt{2}$. $b$ – The result of the superposition of these two beams: a cross-section through the line $ab$ of the distribution of intensity of the CB, $I_c(x,y)$; for the point of maximal intensity, $a$, $r_a \approx 0.9 cm$, for point $b$, $r_b \approx 0.71 cm$. $c$- Contours of the rotating distribution, $I_c(x,y,t)$, at an instant of time. In the colored areas, the intensity, $I_c(x,y,t) > 0.5 \cdot I_{c,\max}(x,y,t)$.

The laser beam described by (13) is remarkable in the sense that its wave front is a helical surface with field intensity equal to zero at the axis of propagation, $z$. In any transverse plane, the field intensity is distributed according to an axially symmetric light circle. The dark central area is caused by the initial helical phase perturbation, $\varphi(r) = \beta$. As a result of this perturbation, the "current streamlets" of electromagnetic energy are the helical lines, which enclose the vortex axis.

The parameters of the beams (the amplitudes, $A$ and $B$, and the radii, $r_{v0}$ and $r_{g0}$) are chosen so that the distribution of the intensity of the Gaussian beam, $I_g(x, y, z = 0)$, is imbedded in the "funnel" provided by the distribution of intensity of the OV, $I_v(x, y, z = 0)$. (See Fig. 5a). In this case, there is a circle with radius, $r_b$, on which $|U_{ov}(r_b)| = |U_g(r_b)|$. As the phase of the OV varies from $0$ to $2\pi$ on this circle, a point, $b$, exists at which the phases of two beams opposing and $U_c(r_b, \beta, z = 0) \equiv U_{ov}(r_b, \beta, z = 0) + U_g(r_b, \beta, z = 0) = 0$. At the radially opposite point, $a$, the intensity of the combined beam (CB), $I_c(r_a, \beta + \pi, z = 0) = |U_c(r_a, \beta + \pi, z = 0)|^2$, is maximal. The result of the superposition of the two beams at time $t = 0$ is shown in Fig. 5b.

We consider here the case for which the frequencies, $\omega_{ov}$ and $\omega_g$, differ by the amount, $\delta\omega = |\omega_{ov} - \omega_g| \ll \omega_{ov}, \omega_g$. In this case, the asymmetric distribution, $I_c(x, y, z = 0)$, rotates around the $z$-axis with the frequency, $\delta\omega$, anticlockwise (if $\omega_{ov} > \omega_g$): the maximum of the intensity occurs at $r = r_a$, with the azimuthal angle, $\beta_a = const - \delta\omega \cdot t$. The value, $\delta\omega$, can be chosen in the desired region (e.g. $\delta\omega \sim 10^{10}\, Hz$ and above).

In a *homogeneous atmosphere,* the intensity maximum of the CB and its center, $\left(\vec{r}_c(z) = \int I_c(x, y, z)\vec{r}ds \big/ \int I_c(x, y, z)ds\right)$, deviate from the $z$-axis due to diffraction spreading. In the plane of the detector perpendicular to the $z$-axis, both points move on concentric circumferences with radii of $r = r_a(z)$ and $r_c(z)$ ($r_a > r_c$). Thus, a CB is a continuous series of PCBs, oriented in the direction of detector with scattering parameter, $\rho(z) \equiv r_a(z)$. (From two characteristic values, we have chosen the more physically visible.)

As shown above, in a turbulent atmosphere, this set of PCBs can significantly compensate the wandering. This leads to a reduced SI for the averaged intensity,

$$\hat{I}^l = \frac{1}{2\pi}\int_0^{2\pi} I^l_{\beta_a} d\beta_a. \qquad (14)$$

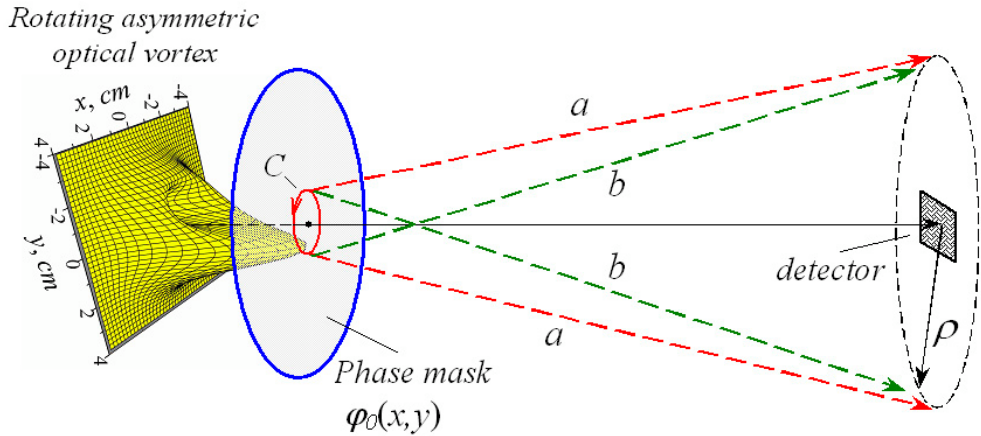

Figure 6. The scheme of formation of a continuous set of PCBs created by a rotating asymmetrical optical vortex.

In (14) the averaging was done over one period of rotation of the CB, $0 \leq \beta_a \leq 2\pi$. $\beta_a$ is the azimuth of the maximum of $I_c(x, y, z = 0)$.

For useful suppression of the SI, we used an additional and important element in our numerical model–the phase mask, $\varphi_0(x,y,z \approx 0)$, (independent of time. See Fig. 6.) The need for this mask is motivated by the following. For the given parameters of the CB (see Fig. 5) at $z = 10 km$, the scattering parameter is $\rho \equiv r_a(z=D) \approx 16 cm$, and the center of the OV is at $r_b(z=D) \approx 5 cm$. For these conditions, effective compensation of the wandering of the beams with $r_w(z_d = 10 km) \approx 15-35 cm$ ($C_n^2 = 2.5 \times 10^{-14} m^{-2/3}$) cannot be achieved. Consequently, one must position the phase mask in the path of propagation of the CB. This mask can provide beam spreading larger than the diffraction widening. After passing through this mask, the scattering parameter, $\rho_{shot}$, can be increased to the required value at the location of the detector. The optimal choice of the function, $\varphi_0(x,y,z \approx 0)$, is a complicated problem. Averaging (14) over a continuous set of $\beta_a$ is equivalent to averaging over a discrete set $\beta_{a,m}$ ($m = 1, 2, ..., M$) for the proper choice of $M$. The maximal reduction of the SI can be achieved if the values $I_{\beta_{a,m}}^l \equiv I_m^l$ are statistically independent. (See (11)). Consequently the phase modulation, $\varphi_0(x,y,z \approx 0)$, should not increase the region of overlap of the distribution of intensity of the CB, $I_{c,m}(x,y)$ and $I_{c,n}(x,y)$, for different moments of time (for different angles, $\beta_{a,j}$, $j = m, n$). This condition is not satisfied, for example, for the radial modulation, $\varphi_0(r) = -\gamma r^2$.

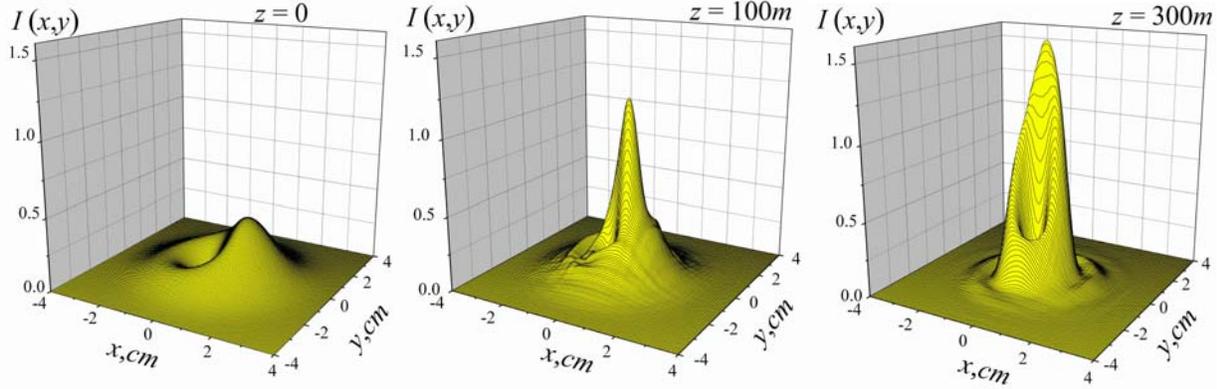

Figure 7. Spatial evolution of the beam after a phase mask-"converging lens" - (15).

Generally, for the given distance, $z_d$, and turbulence strength, $C_n^2$, the maximal possible number $M_{opt}$ (for which statistical independence of the signals $I_m^l(z_d)$, $m = 1, 2, ..., M_{opt}$, is still valid) is determined by the following parameters:

- Radii of the OV and of the Gaussian beams. The diffraction lengths of the beams, (Riley length, $z_{R,ov} = \pi r_{v0}^2 / \lambda$, and $z_{R,g} = \pi r_{g0}^2 / \lambda$), should not be small compared with $z_d$.
- Ratio $r_{v0}/r_{g0}$. Even in the absence of atmospheric turbulence, the combined beam does not remain self-preserving during propagation if $r_{v0}/r_{g0} \neq 1$ (in contrast to the OV and Gaussian beams considered separately).
- The ratio of beam amplitudes, $A/B$, which characterizes a degree of overlapping of PCB intensities in the plane, $z = 0$.
- The form of phase mask, $\varphi_0(x,y)$ (see. Fig.6). In particular, for the same value of $\rho_{shot}$, this mask can simulate a "concave lens" (rays *a* in Fig. 6) or a "converging lens" (rays *b*).

For the combined beam presented in Fig. 5, we have chosen a phase modulation in the form of a cone of revolution

$$\varphi_0(r) = 4\frac{r}{r_{v0}}, \tag{15}$$

in which $r_{v0} = 2cm$ is the radius of the OV. This corresponds to the case of a "converging lens" (rays $b$) in Fig. 6. The transverse cross-section of this beam approaches a minimum at the distances $\sim 300m$ (see Fig. 7), then it diverges, and at $z_d = D = 10km$ $\rho_{shot} \approx 35cm$. The atmospheric turbulence ($C_n^2 = 2.5 \times 10^{-14} m^{-2/3}$) at small distances ($z \leq 300m$) practically did not reveal itself.

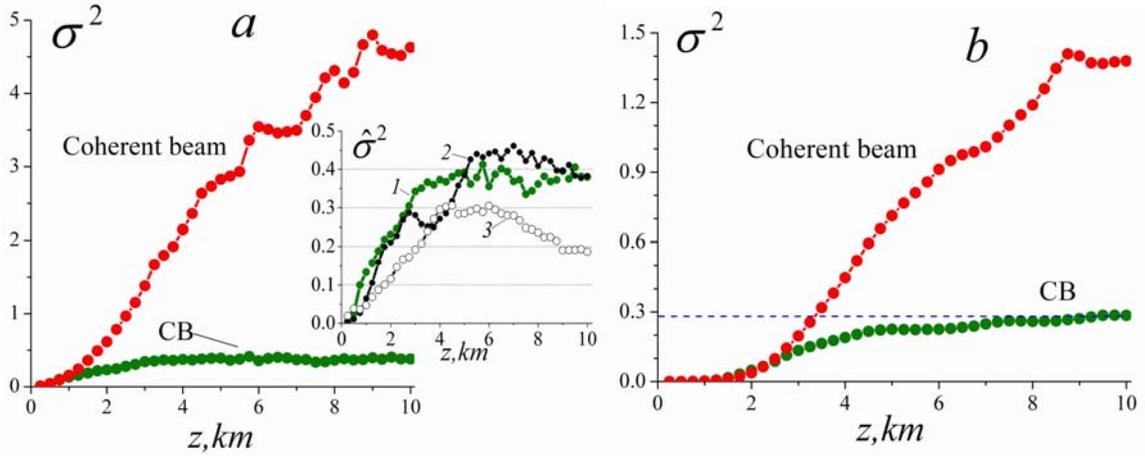

Figure 8. Comparison of SIs calculated for a coherent Gaussian beam ($r_0 = 2cm$) and for a CB ($r_{v0} = 2cm$, $r_{g0} = r_{v0}/\sqrt{2} \approx 1.4cm$, $A = 1$, $B = 0.4$ - see (12)). $a$ -The area of detector $S = 1cm^2$; $b$ - $S = 100cm^2$. Insert in Fig. 8$a$ – a comparison of dependences $\hat{\sigma}^2(z)$ for the combined beam (curve 1), for the coherent beam with a phase mask (9) (curve 2 – the same dependence is shown in Fig. 4), and for the coherent beam with a mask (9) in the regime of monitoring of beam wandering (tracking)- curve 3. In all cases $\rho_{shot} \approx 35cm$, $M = 12$.

The effectiveness of using a combined beam is demonstrated in Fig. 8. Note that the parameter, $\rho_{shot} = 35cm$, for suppression of SI was chosen for $z_d = 10km$. This result is similar to that presented in Fig. 4, except that no high-frequency phase modulator is required! One could assume that a "concave lens", $\varphi_0(r) \approx -4r/r_{v0}$, with the same scattering parameter, $\rho_{shot} \approx 35cm$, is equivalent to the phase mask (15). But the results of numerical simulations demonstrate a significant difference. The effectiveness of a "concave lens" is much less, approximately by a factor 1.5.

Note that a combined beam with an initial helical wave front is more sensitive to the inhomogeneities of the refraction coefficient, $n'$, than a Gaussian beam. The line that connects the deviations of a combined beam $\vec{r}_{w,m}(z_d = D)$ on the $xy$-plane, deviates noticeably away from the circumference and often approaches an ellipse, opposite to the results shown in Fig. 1.

The problem of optimizing the parameters of the combined beam and the phase modulator, $\varphi_0(x,y)$, is quite complicated. The results for the CB presented in Fig. 8 demonstrate an importance similar to those presented in Fig. 4.

However, for the given values of $z_d$ and $C_n^2$, the CB parameters (indicated in Fig. 8) and the phase mask (15) do not correspond to an optimal solution of the problem. For a detector with the area, $S = 1 cm^2$, it is sufficient to choose $M = 6$ (and not 12). For $S = 100 cm^2$, the result shown in Fig. 8b can be achieved for $M = 4$, which indicates the existence of strong signal correlations at the detector, $I_m^l$, if $M = 12$. But this result should be taken with some optimism because this creates opportunities for significantly improving these results using a special time-independent phase modulation, $\varphi_{0,opt}(x,y)$, that is, for example, asymmetric in azimuth $\beta$.

## 4. CONCLUSION

We now summarize our results. These results are essential for the optimal design of high-rate and long-distance optical communications through turbulent atmospheres:

- Significant suppression of the scintillation index (SI) at the detector can be achieved by averaging over a set of states, $m = 1, 2, ..., M$, of a partially coherent beam (PCB). However, the necessary statistical independence of the signals, $I_m^l$, for a given atmospheric state, $l$, required for the method to be effective, is still not guaranteed. Indeed, the turbulent atmosphere traversed by a set of PCBs could produce correlations among the intensities, $\{I_m^l\}_{m=1,2,...,M}$, which decrease the effectiveness of the averaging.

- To reduce the SI, one must compensate (i) for beam wandering and (ii) for the statistical dependence of the signals, $I_m^l$. This compensation can be achieved by using a set of PCBs with preselected angles of propagation relative to the plane of the detector. The phase modulations, $\varphi_m(\vec{r}, z=0) = \vec{a}_m \cdot \vec{r}$, can be used for a predetermined $\{\vec{a}_m\}$ defined in terms of the statistical parameter, $\sqrt{\langle r_w^2 \rangle}$, that characterizes the beam wandering. The use of averaging in this case can produce a significant reduction of the SI. However, the required high-frequency ($\sim 10^9 s^{-1}$) phase modulator is still a major obstacle.

- To overcome the limitations described above, a combined beam (Gaussian and an optical vortex) can be designed to compensate for wandering and to eliminate the need for a high-frequency phase modulator. Designing the corresponding optimal optical system will require additional research to determine the optimal amplitudes and radii for the Gaussian beam and the optical vortex. An important remaining problem is determining the optimal time-independent phase modulation mask, $\varphi_{0,opt}(x,y)$.

## ACKNOWLEDGEMENT

This work was carried out under the auspices of the National Nuclear Security Administration of the U.S. Department of Energy at Los Alamos National Laboratory under Contract No. DE-AC52-06NA25396. The Office of Naval Research also funded this research.